\documentstyle{article}
\newcommand{\bildh}[3]{{\picplace{#2}}\label{#1}}

\newcommand{\thetitle}{}
\newcommand{\thesubtitle}{}
\newcommand{\thedate}{}
\newcommand{\theauthor}{}
\newcommand{\theinstitute}{}

\renewcommand{\title}[1]{\renewcommand{\thetitle}{#1}}

\renewcommand{\date}[1]{\renewcommand{\thedate}{#1}}
\renewcommand{\author}[1]{\renewcommand{\theauthor}{#1}}
\newcommand{\institute}[1]{\renewcommand{\theinstitute}{#1}}

\newcommand{\inst}[1]{$^{#1}$\renewcommand{\and}{, }}
\newcommand{\instn}[1]{$^{#1}$\renewcommand{\and}{\\}}
\newcommand{\thesaurus}[1]{}

\renewcommand{\maketitle}{
\thispagestyle{empty}
{\parindent0cm
\begin{center}
\vskip 2em
{\LARGE \thetitle \par {\Large \thesubtitle}}\par
\vskip 2em {\large \theauthor}\par
\vskip 1em {\theinstitute}\par
\vskip 2em {\thedate}
\end{center}
}}

\newcommand{\la}{\le}

\newcommand{\keywords}{{\bf Keywords:} }

\newcommand{\degr}{\mbox{$^\circ$}}
\newcommand{\picplace}[1]{\frame{\centerline{(will be inserted later)}}}
\newcommand{\acknowledgements}{{\it Acknowledgements.}}
\renewenvironment{thebibliography}[1]{{\section*{References}}
\parindent0cm}{\par\vskip1.5em}
\renewcommand{\bibitem}[3]{\par}

\begin{document}
\thesaurus{01(11.01.2; 11.10.1; 11.14.1; 02.01.2)}
\title{Unified schemes for active galaxies:
a clue from the missing Fanaroff-Riley type I quasar population}
\author{Heino Falcke\inst{1}, Gopal-Krishna\inst{1,2} \and Peter L.
Biermann\inst{1}}
\institute{Max-Planck Institut f\"ur Radioastronomie, Auf dem H\"ugel 69,
D-53121 Bonn, Germany\instn{1} \and National Centre for Radio
Astrophysics (TIFR), Poona University Campus, Pune -- 411007,
India\instn{2}}
\date{Astronomy \& Astrophysics in press [astro-ph/9411106]}
\maketitle
\markboth{Falcke et al.: Unified schemes -- a clue from the missing FR\,I
quasars}{Falcke et al.: Unified schemes -- a clue from the missing FR\,I
quasars}

\begin{abstract}
We link the lack of FR\,I type structure among quasars to the void of
radio loud quasars below a critical disk luminosity of $\sim10^{46}$
erg/sec in the PG sample.  We argue that the opening angle of the
obscuring torus in radio loud quasars depends on the power of the
central engine, approaching the jet's beaming angle near the
FR\,I/FR\,II break. Consequently, low power radio quasars would either
be classified as radio galaxies (FR\,I) or strongly core-boosted (BL
Lac) object, depending on the aspect angle, and no conspicuous
transitional population would be expected for FR\,I sources.  A
closing torus with decreasing power would not only obscure the optical
nucleus for most aspect angles but would also enhance the entrainment
of the cool torus material into the jet stream, causing obscuration
along the jet's periphery, as well as the jet's deceleration to form a
FR\,I source.  Above a critical luminosity, the wider torus allows for
FR\,II type jets and visibility of the nuclear optical emission,
characteristic of radio loud quasars.  Apparently, at the same engine
power the torus opening in radio weak quasars and Seyferts is
substantially wider than in radio loud quasars, probably because of
different dynamics or feeding mechanisms in disk and elliptical
galaxies. This provides a clue for the radio-loud/radio-quiet
dichotomy of quasars if the the jet/torus interaction leads to
injection of relativistic $e^\pm$ via $pp$ collisions. Strong
jet/torus interaction may lead to a substantial injection of secondary
pairs and collimation in radio loud quasars, while weak interaction in
radio weak quasars leads neither to pair injection nor to good
collimation.
\end{abstract}
\keywords{galaxies: active -- galaxies: jets -- galaxies: nuclei -- accretion
disks}

\section {Introduction}
The different manifestations of energetic phenomena (radio jets and
non-stellar optical emission) in active galactic nuclei (AGN) have
provoked attempts to unify several different classes of AGN, by
arguing that a pair of relativistic jets and a coaxial obscuring torus
can yield radically different views of the same object, depending upon
the aspect angle (see reviews by Antonucci 1993; Gopal-Krishna 1994).
Observations by Lawrence \& Elvis (1982), Mushotzky (1982) and
Antonucci \& Miller (1985), suggested that Seyfert 2 galaxies are
simply Seyfert 1 galaxies where an obscuring torus blocks a direct
view of the nuclear region. Several authors (e.g. Barthel 1989)
extended this and identified the powerful narrow-line radio
galaxies of Fanraroff-Riley type II (FR\,II, edge brightened) as the
misaligned parent population of radio loud quasars, while the
core-dominated, rapidly variable quasars ('blazars') represent the
cases of close alignment between the jet and the
line-of-sight. Following Blandford \& Rees (1978), BL Lac objects are
identified as the boosted counterparts of intrinsically weaker
Fanaroff-Riley type I (FR\,I, edge darkened) radio galaxies (e.g. Urry
et al. 1991).

Despite the consensus that both FR\,I and FR\,II galaxies are
associated with massive ellipticals, it has been argued that they are
distinct phenomena (e.g., Heckman et al.  1994) and hence, the unified
schemes for FR\,I and FR\,II sources would bear no direct
relationship.  On the other hand, it has also been suggested that the
two morphological types may have a physical, perhaps evolutionary
connection (e.g., Owen \& Ledlow 1994).  Also, it has been recently
argued that AGN of all radio powers, including those as weak as the
Galactic Center source Sgr A* (Falcke et al. 1993a\&b, Falcke 1994a\&b),
have basically similar central engines, i.e. a closely coupled
jet/disk system around a black hole, and the apparent differences are
caused largely due to interaction with the parsec-scale environment
(Falcke \& Biermann 1994 (FB94); Falcke, Malkan, Biermann 1994 (FMB)).

Clearly, finding a common ground between the two unification schemes
(FR\,I radio galaxy-BL Lac and FR\,II radio galaxy-quasar-blazar)
would be a consolidating step forward.  One observation that stands in
the way of such a reconciliation is the persistent lack of unambiguous
examples of FR\,I type quasars.  Here we examine this point closely,
using a well defined sample of optically selected quasars and propose
that a {\em closing torus}, rather than fundamentally different types
of engines might be responsible for the striking rarity of FR\,I
quasars {\it and} for the FR\,II/FR\,I transition.

\section{The 'void' of radio-loud quasars}
Although the basis of the unified schemes proposed for FR\,I and
FR\,II radio sources is essentially the same relativistic jet
phenomenon in the nucleus, the two schemes differ in that, unlike the
FR\,II scheme,
the FR\,I scheme does not have to contend
with a 'transitional population' between the parent population (radio
galaxies) and its strongly Doppler-boosted subset (BL Lacs). From
radio imaging surveys, this point has been noted in the past (e.g., Perlman
\& Stocke 1993) and can be perceived more clearly by considering the
radio morphological content of a large sample of optically-selected
quasars, e.g. the PG quasar sample (Schmidt \& Green 1983). In one
such study using the VLA (A-array), Miller et al. (1993) investigated
the radio morphologies of the $z<0.5$ quasars in the PG sample. Out of
the total 89 quasars 13 can be classified as radio-loud and all of
them resemble a FR\,II morphology, consistent with their high radio
luminosities.  As seen from their Fig. 2, the total luminosities of the
13 radio-loud quasars are clustered in a narrow range between
$10^{32}$ and $10^{33}$ erg/sec/Hz/ster at 5 GHz.  Below this limit,
for two decades in luminosity, a conspicuous gap is present in which
not a single radio loud quasar possessing resolved radio emission
(potentially FR\,I morphology) is found. A sharp drop in the
radio-loud fraction of quasars has also been noted to occur below $M_{\rm
B} \sim -24.5$ (Padovani 1993).

This pattern is clearly illustrated in Fig. 1 which shows for the same $z<0.5$
subset of the PG sample a plot of radio {\em core} luminosity
vs. disk luminosity, $L_{\rm disk}$. To estimate the latter,
we combined accretion disk fits,
emission line luminosities and continuum fluxes and obtained an average
 value for each quasar (FMB). Although
the region below a critical $L_{\rm disk}
\la 10^{46}$ erg/s contains 68 quasars, not a single example of a
typical, lobe-dominated radio source is found there. Thus, intrinsic
weakness of the nuclear ionizing radiation, or any related selection
effect can not be the reason for this void.  The existence of a
'critical' $L_{\rm disk}$ is, in fact, reminiscent of the 'break' in
the radio luminosity function of E/SO galaxies, below which powerful
radio sources (FR\,II) become rare and the weaker FR\,I type sources
begin to predominate (Fanaroff \& Riley 1974) -- in fact the {\em
total} power at 5 GHz of the weakest FR\,II PG quasar
($2\cdot10^{42}$erg/sec) is very close to the FR\,II/FR\,I break at
$\sim10^{42}$erg/sec.  The morphological transformation at this break
was proposed to be due to the effect of the hot gaseous halo of the
parent elliptical galaxy on the advancing jet. Below the critical
power, the external medium decelerates the jet at an early stage to
subsonic velocities, leading to a rapid decollimation and enhanced
mass entrainment, all resulting in radio fading and FR\,I morphology
(Gopal-Krishna 1991; Roland et al.  1992; Bridle 1992). Jets above the
critical power would continue to propagate supersonically and
terminate far away in a strong shock producing compact hot spots.  In
this picture, the jets responsible for the two morphological types are
not required to be basically different {\it ab initio}.

According to the above scenario for FR\,II to FR\,I transition with
decreasing source power, and considering that the jet and the UV bump
are both produced by the same engine (e.g. FMB), one might naively
expect a similar transition to occur also in the case of quasars: a
lower accretion power would presumably yield both a lower disk
luminosity and a weaker jet (FB94) and thus below a critical disk
luminosity of $L_{\rm disk}\la10^{46}$ erg/sec FR\,I quasars should be
found.  Apparently, this does not happen; the disappearance of FR\,II
quasars below the critical disk luminosity is not accompanied by an
emergence of a FR\,I quasar population (Fig.  1). It may be noted that
the appearance of the latter in Fig. 1 would, in fact, be facilitated
by the fact that compared to FR\,II sources, the radio cores in FR\,I
sources are relatively more prominent (Bridle 1992). It can therefore
be surmised that {\it the void of radio-loud quasars below $L_{\rm
disk} \la 10^{46}$ erg/sec is in fact the void of FR\,I quasars}.

\begin{figure}
\centerline{\bildh{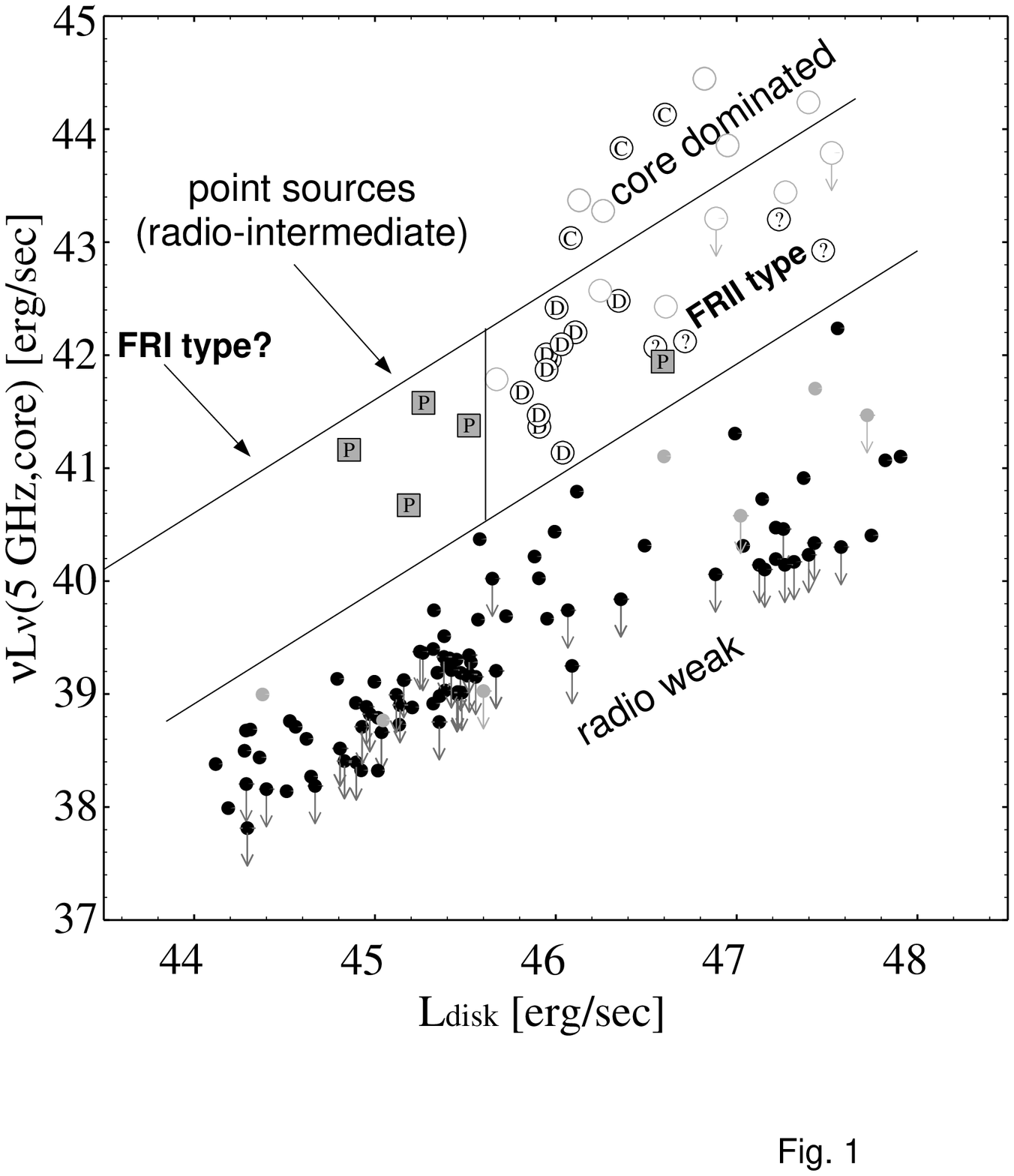}{8cm}{bbllx=4.7cm,bblly=17.5cm,bburx=13cm,bbury=25.8cm}}
\caption[]{Radio core luminosity vs. the
disk luminosity of PG quasars. Below $L_{\rm disk}\sim10^{46}$ erg/sec
no radio loud quasars of FR\,II type are found (see FMB for details)
-- {\em open circles}: radio loud quasars, `D' marks FR\,II morphology
and `C' marks core dominated quasars; {\em dots}: radio weak quasars;
{\em boxes}: variable flat spectrum points sources, interpreted as
boosted radio weak quasars; {\em grey symbols:} non-PG quasars from
the extended sample of FMB. }
\end{figure}

\section{Approach to harmonise the two unified schemes}
Does the lack of FR\,I quasars, in contrast to the abundance of FR\,I
galaxies, imply that they are different types of systems?  Or,
alternatively, do FR\,I quasars also exist, but are not classified as
such?  Here we argue for the latter by examining the possible
connection between the power of the central engine/jet and the
obscuring torus. From the optical identification content of the 3CRR
complete sample of radio sources, a systematic increase in the
fraction $f$ of the broad-lined objects and, therefore, in the opening
angle of the putative torus with radio luminosity can be inferred
(Lawrence 1991).  The trend appears to be systematic and with a
reasonably high statistical significance --- though reclassification
of some objects may eventually change the numbers
slightly. Furthermore, a corraborative evidence for this trend comes
from the optical identification content available for yet another
complete sample of radio sources, namely, the 1-Jy sample defined by
Allington-Smith (1982; see Singal 1993). It can thus be inferred that,
on average, the bi-conical openings of the torus shrink steadily with
decreasing power of the central engine. From Fig. 2b of Lawrence
(1991), the half-opening angle $(\psi)$ is estimated to be typically
in the range 45\degr{} to 60\degr{} for powerful (FR\,II) sources.
Further, although the observed decrease in $f$ to $\la0.1$ near the
FR\,I/FR\,II break still allows for a sig\-nificant population of
broad-lined objects even at these modest luminosities, it translates
to a $\psi$ of $\la 25\degr$ which with decreasing power below the
break approaches quickly the typical beaming angle of the jets.
Consequently, a direct view of the nuclear region in moderately
powerful (FR\,I) radio sources -- a pre-requisite for quasar
classification -- would be unavoidably accompanied by a strongly
Doppler-boosted appearance of the synchrotron jet. Due to this and the
enhanced obscuration of the nuclear region caused by the jet-
entrained torus material (Sect. 4), the aligned FR\,I sources would
generally end up being classified as BL Lacs.  Cases of an
intermediate jet-alignment would indeed appear less core-dominated but
the nuclear region would be directly obscured by the torus with narrow
openings at these modest powers, leading to a (FR\,I) galaxy
classification.  The lack of a transitional, FR\,I quasar population
can thus be understood (Gopal-Krishna 1994) without invoking any
conceptual difference between the unified schemes for FR\,I and
FR\,II.

\section{Jet interaction with a closing torus}
According to the current understanding, continually accreted molecular
clouds with a wide range in size form a major constituent of the dusty
torus (Krolik \& Begelman 1988). Being directly irradiated by the
energetic photons from the central engine, the inner walls of the
torus are steadily stripped, forming an outflowing wind filling the
torus openings (e.g., Balsara \& Krolik 1993). In highly luminous
sources, the wind filling the funnels is likely to consist of hot
plasma whose thermal pressure could even contribute substantially to
the jet's confinement on the parsec scale as required in some models
(e.g.  Appl \& Camenzind 1993).  The interaction between the jet and
the dense torus material would be more direct in lower power sources
with narrower torus openings -- especially when $\psi$ approaches the
jet's opening angle. The material stripped from the torus walls and
entrained into the jet flow would then rapidly decelerate it to low
Mach numbers and a trans-sonic FR\,I type flow, leading eventually to
the jet's disruption and consequent fading of the radio lobes, as
discussed in the models of Bicknell (1994) and De Young (1993).  As
the entrained clouds would tend to occupy the peripheral region of the
jet's cross section, the boosted nonthermal radiation from the jet's
core would escape essentially unobscured by the clouds.  However, the
optical/UV emission from the outer regions of the jet and especially
from the nuclear disk surrounding the jet near its base (into which
the broad- line-emitting clouds seem to be concentrated; cf. Wills \&
Browne 1986) would probably be strongly obscured.  Thus, a closing
torus with decreasing engine power could cause the obscuration of the
nuclear optical line and UV emission, {\it as well as} the transition from
a FR\,II to FR\,I morphology. This could explain why the radio
morphological transition occurs at a power similar to where the radio
loud quasars practically disappear (Fig. 1). Further, the existence of
a critical optical/radio power for the FR\,I/FR\,II break indicates
that the power-dependence of $\psi$ is probably linked to the
interaction with the jet. A stronger jet-torus interaction could
well be responsible for the observational result that in radio
galaxies the jet appears systematically more prominent with decreasing
total radio power (Bridle 1992).

Here, we mention a specific mechanism by which a locally enhanced
synchrotron emissivity could arise during a strong interaction of the
jet with a dense torus. It was argued from synchrotron and Compton
loss arguments (e.g. FB94) that in order to explain the high
efficiency of radio loud jets in a coupled jet/disk system additional
relativistic electrons have to be injected at the $0.1-1$ pc scale.  A
possible mechanism for the radiation enhancement would be $e^{\pm}$
pair production in hadronic cascades following $pp$ collisions of
relativistic protons (Biermann \& Strittmatter 1987; Sikora et al.
1987) with fragments of the dense material in the torus funnel or in
the shear layer between the jet and the ambient medium. Although
$e^{\pm}$ pairs in the jets are generally thought to arise primarily
from electromagnetic processes near the center, the
possibility of a substantial contribution from  hadronic
cascades has been defended recently on energetic grounds, since such
collisions would naturally yield a low energy cut-off around $50-100$
MeV (FB94, Biermann et al. 1994), as inferred from independent
arguments (Wardle 1977, Celotti \& Fabian 1993).

\section{Radio-weak quasars}
While the FR\,I/FR\,II radio galaxies and radio-loud quasars are
almost exclusively hosted by ellipticals, most radio-quiet quasars and
Seyfert nuclei appear to reside in disk galaxies (e.g. McLeod \& Rieke
1994). The small number of luminous Seyfert 2 (quasar 2) (Osterbrock
1991), the rarity of highly polarized Seyfert 2 (Miller
\& Goodrich 1990) and the evidence for a declining covering factor of the
central engine with increasing luminosity (e.g., Lawrence \& Elvis
1982; Reichert et al. 1985) have led to the speculation that the
torus-opening in Seyferts and quasars becomes wider with
increasing power of the engine. From Sect. 2 and 3, we inferred that
even at a disk luminosity as high as $10^{46}$ erg/sec, the torus in
radio-loud AGN is practically closed, whereas it is already wide open
in the case of radio-weak AGN ($\psi$ approaching 60\degr; see, McLeod
\& Rieke 1994; Dunlop et al. 1993).  Even in exceedingly radio-weak
AGN, like Seyferts, the typical value of $\psi$ is still quite large
(about $30\degr$, e.g. Pogge 1989).  From this we infer that for a
given power of the central engine, a radio-weak AGN can maintain a
wider torus opening.  This is perhaps a result of higher specific
angular momentum expected for the nuclear tori within their host
(disk) galaxies. Since ellipticals with AGN are thought to be merger
products (e.g., Barnes \& Hernquist 1993), a significant cancellation
of the angular momentum of the progenitors can occur, in contrast to
spirals where the gas is probably accreted from the disk, stirred up
by starbursts (Biermann et al. 1993).  Camenzind (1993) has suggested
that tori with narrow opening angles could be due to the substantially
larger core radii in ellipticals.  It is tempting to speculate that a
wider torus opening in disk galaxies leads to a poor confinement of
the jet and a further loss of radio emissivity via strongly diminished
injection of secondary pairs in hadronic cascades due to a weaker
jet-torus interaction and less $pp$ collisions. A poorly confined jet
ejected from a disk would be slowed down much earlier, producing
diffuse, radio-inefficient lobes in radio weak quasars (Miller et
al. 1993).  Thus, a different type of torus in disk galaxies could be
a significant factor contributing to their radio weakness.

\section {Conclusions}
 Using the PG sample of QSOs, we have argued that the near absence of
 quasars with FR\,I type radio morphology could be a direct
 consequence of the reported trend for the torus to be wider open in
 more luminous radio sources. Near a critical disk luminosity ($\sim
 10^{46}$ erg/s), the torus narrows down to roughly the beaming angle
 of the relativistic jet ($10\degr$ to $15\degr$), inhibiting a direct
 view of the nuclear region without the appearance of a strongly
 Doppler-boosted (BL Lac like) synchrotron jet. Moreover, below this
 power, a further shrinking of the funnels would cause a direct
 interaction of the jet with the torus, triggering the transition from
 an FR\,II to FR\,I morphology. This transformation would be
 accompanied by an increased obscuration of the optical line and
 continuum emission from the nuclear disk by the dense torus material
 entrained in the jet flow. Collisions of relativistic protons in the
 boundary layer between jet and torus funnel could locally enhance the
 synchrotron output of the jet and, hence, its prominence by injecting
 secondary pairs in hadronic cascades (Fig. 2). Consequently, there
 seems to be no need for conceptually different unification schemes
 for FR\,I and FR\,II.
\begin{figure}
\centerline{\bildh{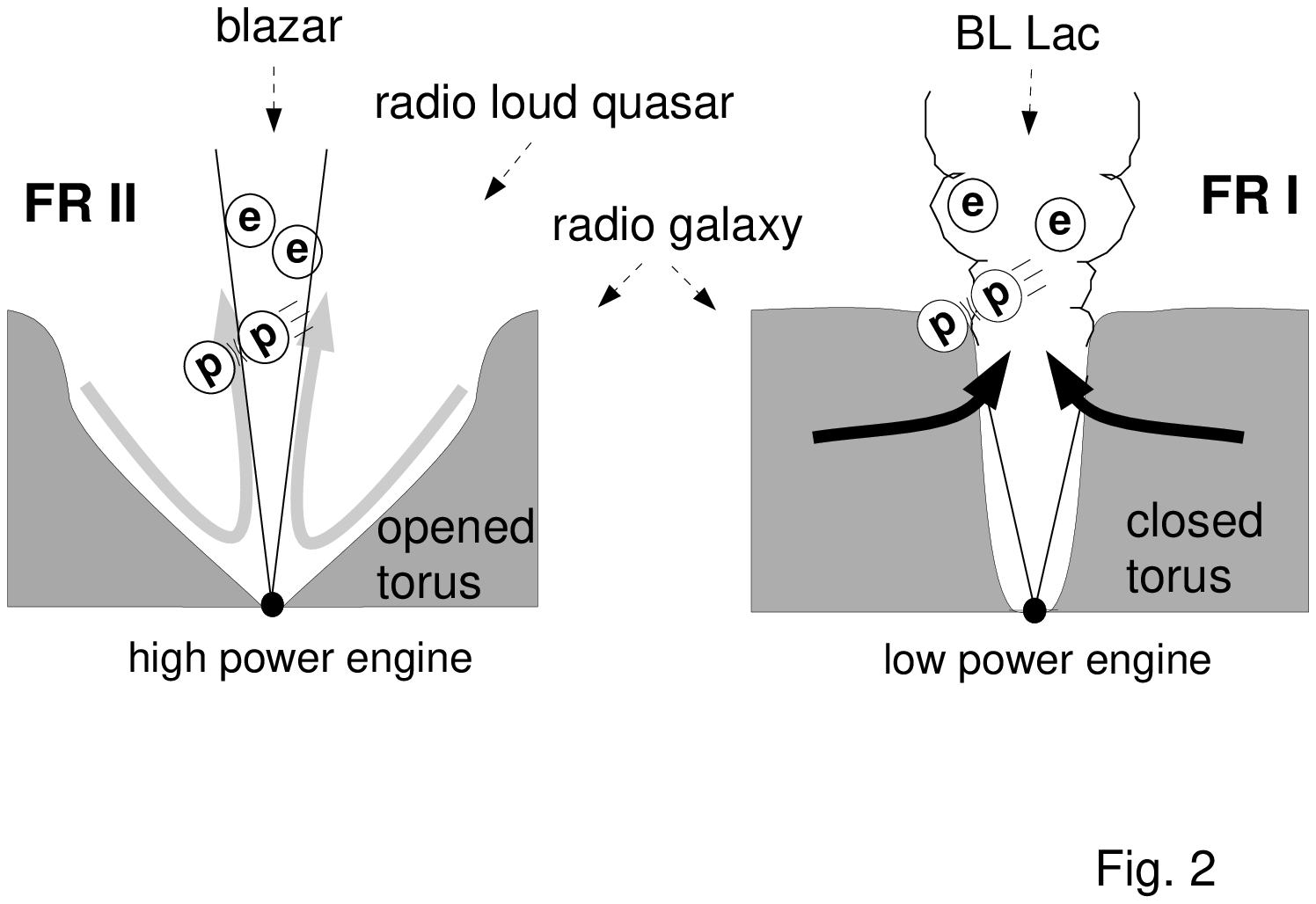}{5cm}{bbllx=4.1cm,bblly=20.8cm,bburx=11.6cm,bbury=24.5cm}}
\caption[]{Sketch of the proposed scheme: a low-power (FR\,I) jet
is slowed down by a closing torus which also obscures the central
engine while a high-power (FR\,II) source with open torus appears as a
quasar for some aspect angles. Hadronic cascades might produce
additional relativistic pairs in the interaction zone.}
\end{figure}

The critical disk power of $\sim L_{\rm disk}\sim10^{46}$ erg/s in
ellipticals where the torus appears closed apparently is a few
orders-of-magnitude higher than the corresponding value for
spirals. Thus, the intrinsically wider torus at the same
engine power in spirals would result in a fainter jet due to poor
confinement and weaker jet-torus interaction causing a diminished
injection of secondary $e^\pm$ into the jet. It appears plausible that
the type of the host galaxy is more important in determining the shape
of the torus, rather than the much smaller region dominated
by the black hole. Hence, {\it intrinsically different tori at the pc
scale, rather than intrinsically different central engines} might play
a critical r\^ole for the radio loud/radio-weak quasar dichotomy and
the FR\,I/FR\,II separation.

A key question for future work would be the formation and stability  of
tori in different types of host galaxies. Papaloizou \& Pringle (1985)
found that thick accretion disks have unstable modes, which, however,
depend on the boundary conditions at the edge of the disk (Goldreich et
al. 1986) -- a powerful jet could strongly influence this result.  The
torus could also fragment into clouds (Hawley 1987), yielding a configuration
similar to the Krolik \& Begelman (1988) torus composed of molecular
clouds, but the stability and structure of the latter for very small
$\psi$ remain unclear. The reason for the power dependence of $\psi$
could well be the jet/torus interaction but also radiation
pressure or a changing disk structure with accretion rate.

Further observational consequences are expected in the high energy
regime, i.e.  production of X-rays and gamma-rays (see also Mannheim
1993) and in the infrared (IR). Since the torus in FR\,II already
intercepts more than half of the central luminosity, only a marginal
increase in the fraction of nuclear radiation reprocessed into the IR
is expected for FR\,I.  However, due to the different shapes of their
torus-funnels, spectral differences in the IR are possible.  One also
expects to find the ionization cone associated with FR\,I aources to
be quite narrow.  In case the radio weak quasars, too, eject
relativistic jets one ought to find their boosted counterparts (RIQs,
see FMB).

\acknowledgements
HF thanks L. Saripalli for stimulating discussions on the puzzling
lack of FR\,I quasars. We are grateful to the referee Dr. M. Salvati
for constructive criticism. We acknowledge additional, delayed reports
by another referee, persistently reminding us of the fact that the
standard unification schemes quoted by us are still rejected by some
workers in -- and outside -- the field.

\end{document}